\input epsf
\documentstyle[epsfig,eqsecnum,aps,prd,preprint,tighten,floats]{revtex}
\begin{document}
\newcommand{\gras}[1]{\mbox{\boldmath $#1$}}
\newcommand{\be}{\begin{equation}}
\newcommand{\ee}{\end{equation}}
\newcommand{\ba}{\begin{eqnarray}}
\newcommand{\ea}{\end{eqnarray}}
\newcommand{\Mc}{{\cal M}}
\newcommand{\Ms}{M_{\odot}}
\newcommand{\m}{\langle}
\newcommand{\M}{\rangle}
\newcommand{\hf}{\tilde h}
\newcommand{\rf}{\tilde r}
\newcommand{\yf}{\tilde y}
\newcommand{\qf}{\tilde q}
\newcommand{\Qf}{\tilde Q}
\newcommand{\bml}{\begin{mathletters}}
\newcommand{\eml}{\end{mathletters}}
%
%minore o circa uguale
\def\laq{~\raise 0.4ex\hbox{$<$}\kern -0.8em\lower 0.62
ex\hbox{$\sim$}~}
%maggiore o circa uguale
\def\gaq{~\raise 0.4ex\hbox{$>$}\kern -0.7em\lower 0.62
ex\hbox{$\sim$}~}

\preprint{\vbox{\baselineskip=12pt
\rightline{BA-TH/01-412}}}
\rightline{BA-TH/01-412}
\preprint{\vbox{\baselineskip=12pt
\rightline{RCG 01/10}}}
\rightline{RCG 01/10}

\rightline{gr-qc/0103035}

\vskip 2true cm

\vspace{10mm}
{\large\bf\centering\ignorespaces
Detecting a relic background of scalar waves with LIGO\\
\vskip2.5pt}

\bigskip
{\dimen0=-\prevdepth \advance\dimen0 by23pt
\nointerlineskip \rm\centering
\vrule height\dimen0 width0pt\relax\ignorespaces
 M. Gasperini${}^{(1,2)}$ 
 and C. Ungarelli${}^{(3)}$
\par}
\bigskip
{\small\it\centering\ignorespaces
${}^{(1)}$
Dipartimento di Fisica, Universit\`a di Bari, \\
Via G. Amendola 173, 70126 Bari, Italy\\
${}^{(2)}$
Istituto Nazionale di Fisica Nucleare, Sezione di Bari,
Bari, Italy \\
${}^{(3)}$
Relativity and Cosmology Group\\
School of Computer Science and Mathematics,
University of Portsmouth,\\
Portsmouth P01 2EG, England
\par}

\par
\bgroup
\leftskip=0.10753\textwidth \rightskip\leftskip
\dimen0=-\prevdepth \advance\dimen0 by17.5pt \nointerlineskip
\small\vrule width 0pt height\dimen0 \relax

\begin{abstract}
We discuss the possible detection of a stochastic background of massive,
non-relativistic scalar particles, through the cross correlation of the two
LIGO  interferometers in the initial, enhanced and advanced
configuration. If the frequency corresponding to the mass of the scalar
field  lies in the detector sensitivity band, and the non-relativistic branch
of the  spectrum gives a significant contribution to energy density
required to close  the Universe, we find that the scalar background can
induce a non-negligible signal, in competition with a possible signal
produced by a  stochastic background of gravitational radiation. 
\end{abstract}

 \par\egroup
\thispagestyle{plain}

\pacs{}

\section{Introduction}
\label{S1}

In the next few years many gravitational antennas will be collecting
data,  as the new interferometric detectors
(GEO, LIGO, TAMA, VIRGO) \cite{1} will join the resonant detectors
(ALLEGRO, AURIGA, EXPLORER, NAUTILUS, NIOBE) \cite{2} already in
operation, covering a frequency band from $\sim 10$~Hz up to $1$KHz. 
Through the cross-correlation of these antennas it will be
possible to search for  a stochastic background of
primordial gravitational radiation \cite{3}, \cite{3a}.
The detection of  a relic background of gravitational 
waves would allow us  to reconstruct the very early stages of 
cosmological evolution~\cite{4} and even an upper limit would be 
significant, since there are predictions of very high relic 
backgrounds \cite{5} in principle detectable already by second 
generation interferometers \cite{6} (see however \cite{6a}). 

A generic prediction of unified theories (such as supergravity and 
superstring theories) is the existence of a gravitational multiplet which 
includes, beside the usual spin-2 graviton, scalar components. In a 
cosmological context, as well as the production of a primordial 
background  of gravitational waves, it is therefore worth investigating
the possible  production of a relic background of scalar waves (for
instance, dilatons \cite{7}). In particular, if the mass of the scalar field is
small  enough ($ m \laq 100$ MeV), such a background would be present
even today, and it could be accessible to direct experimental
observation. It seems thus appropriate to estimate the sensitivity of
gravitational antennas also to scalar waves. 

A relic background of scalar particles can interact with a gravitational
antenna in two ways: either indirectly, through the geodesic coupling
of the detector to the induced background of scalar metric
fluctuations, or even directly \cite{8}, through the effective scalar
charge of the detector (such a direct coupling cannot be absorbed into
the metric interactions, in an appropriate frame, if the scalar
charge is non-universal). Up to now, the sensitivity to scalar waves has
been mainly studied using the indirect coupling of the antennas to the
scalar part of metric fluctuations, and considering massless scalar fields
(see for instance \cite{9}). Under those assumptions,  the only difference
with respect to the case of  standard gravitational radiation is
represented by  the polarisation tensor of the scalar wave~\cite{10}. 

However, if the detector response tensor is characterised by a 
symmetric  $3\times 3$ trace-free tensor (such as the differential mode
of an  interferometer), the sensitivity to a scalar wave of momentum $p$
and energy  $E$ generated by a massive scalar field is suppressed 
by a factor~\cite{10} $(p/E)^4$ with respect to the sensitivity to an 
ordinary gravitational wave\footnote{This  factor coincides 
with the suppression factor for the cross section of a
resonant bar to longitudinal massive scalar modes \cite{8}.}. This 
suppression is ineffective only for modes which are
ultrarelativistic in the sensitivity band of the detector, 
$m \ll E= (p^2+m^2)^{1/2} \simeq p$; this condition, however, occurs
when the frequency corresponding to the mass of the scalar field is much
smaller than the typical frequency  of the detector, i.e. when $m \ll E_0
\sim 10 -10^3$ Hz.  The  scalar field would be  then associated to
long-range interactions, and  its  coupling to the mass of the detector
should be  highly suppressed (with respect to the standard gravitational
coupling), in order to agree with the existing tests of the equivalence
principle and of macroscopic gravitational interactions \cite{11}.  
Hence, as far as interferometric antennas are concerned, the possible 
detection of a background of (massive or massless) scalar waves would
seem to be strongly unfavoured with respect to the detection of a
graviton background  (in the 
literature,  indeed, the possible detection of scalar waves is usually
demanded to future resonant detectors of spherical type \cite{12}). 

Nevertheless, it is important to notice that, when the mass of the scalar 
field corresponds to a frequency which is in the detector sensitivity 
band,  it is possible to obtain a resonant response also from the
non-relativistic part of the scalar waves spectrum \cite{13}.  On the 
other hand, unlike a relativistic background of massless particles (like 
gravitons),  the non-relativistic part of a relic background is not
constrained by the nucleosynthesis bound, and could saturate  the critical
energy density  required to close the Universe. As 
suggested in in \cite{13}, if the non-relativistic branch of the spectrum is 
peaked at  $p \sim m$, the polarisation suppression factor and the
weakness of the scalar coupling could be compensated by the high relic
density, and such a background of scalar waves could be a 
potential source for interferometric detectors.

The aim of this paper is to discuss the possible detection
of a relic stochastic background of massive scalar particles with
the two LIGO interferometers (including the enhanced and advanced
configurations). The paper is organised as follows. In Section \ref{S2} we
recall the general expression for the optimised signal-to-noise ratio (SNR), obtained by cross-correlating two detectors, with respect to a stochastic 
background of massive scalar waves. In Section
\ref{S3} we apply this result to the case of the 
differential mode of an interferometer, 
taking into account both the geodesic coupling to the
scalar part of the metric fluctuations, and the direct coupling to the
scalar field. In Section \ref{S4} we discuss some examples, and
we show in particular that a relic background of non-relativistic scalar 
particles, whose energy density provides a significant fraction of 
the critical energy density,  can induce 
a non-negligible signal in the cross-correlation of the LIGO 
interferometers (in the enhanced and advanced configurations), if  
the frequency corresponding to the mass of the scalar field is in the
sensitivity band  of the detectors. The main results of this paper are
finally summarised in Section \ref{S5}.  

\section{Signal-to-noise ratio}
\label{S2}

We will consider a  stochastic background of massive
scalar particles, described in terms of a scalar field 
 $\phi(\vec{x},t)$, and characterised by a dimensionless
spectrum $\Omega(p)$, 
\be
\Omega(p)= {1\over \rho_c} {d \rho\over d \ln p},~~~~~~~~~~~~
\rho_c={3H_0M_P^2\over 8 \pi},
\label{21}
\ee
where $p$ is the momentum, $\rho$ is the scalar field energy density, 
$\rho_c$ is the critical energy density, $H_0$ the present value
of the Hubble parameter, and $M_P$ the Planck mass. We shall assume
that the scalar field is coupled to the mass of the detector with
gravitational strength (or weaker), and that the spectrum extends in
momentum space from $p_0$ to $p_1$. The low frequency cut-off
$p_0$ may be zero (for growing spectra), or the present Hubble scale
(for decreasing spectra), while $p_1$ is an high-frequency cut-off which
depends on the details of the production mechanism. 
 
Our starting point is the expansion of the scalar field  in the 
momentum space, 
\be
\phi(t,\vec{x})=\int_{0}^{\infty}dp\,
\int\,d^2\hat{n}\left\{
e^{2\pi\,i[E(p)t-p\hat{n}\cdot\vec{x}]}{\phi}(p,\hat{n})
+\mbox{h.c.}\right\}\,,
\label{decomp}
\ee 
where $\hat{n}$ is a unit vector specifying the propagation direction, 
$\vec{p}=\hat{n}\,p$ is the momentum  vector,  
and the energy $E(p)$ of each mode is 
\be
E(p)=\sqrt{p^2+(m/2\pi)^2}\,   
\ee
(we are adopting ``unconventional" units $h=1$, so that the
proper frequency is simply $f=E$). The 
background of scalar waves is assumed to be isotropic, stationary
and Gaussian~\cite{3} and satisfies the following stochastic conditions 
\ba
&&\langle{\phi}(p,\hat{n})\rangle=0\,, \nonumber\\
&&\langle{\phi}(p,\hat{n})
{\phi}^*(p',\hat{n}')\rangle=
\delta(p-p')\delta^2(\hat{n}-\hat{n}')\Phi (p)\,,
\label{stoc_prop}
\ea
where, using the explicit definition of $\Omega(p)$,  
\be
\Phi (p)=\frac{3\,H^2_0\Omega(p)}{8\pi^3\,p\,E^2(p)}\,.
\ee 
The scalar background induces on the output 
of the gravitational detector a strain
$h_{\phi}(t)$, proportional to the value of the scalar field 
$\phi(t,\vec{x}_D)$ at the detector 
position \cite{10}, 
\be
h_{\phi}(t)=\int_{0}^{\infty}dp\,
\int\,d^2\hat{n},F_{\phi}(\hat{n})\left\{
e^{2\pi\,i[E(p)t-p\hat{n}\cdot\vec{x}_D]}{\phi}(p,\hat{n})
+\mbox{h.c.}\right\}\,,
\label{26}
\ee
where $\vec{x}_D$ is the position of the detector centre of mass, and 
$F_{\phi}$ is the antenna pattern. In particular, 
\be
F_{\phi}(\hat n)=q e_{ab}\,D^{ab}\,
\label{27}
\ee
where $e_{ab}$ is the polarisation tensor of the scalar wave,  $D_{ab}$ is
the detector  response tensor, and $q$ is the effective coupling
strength of the scalar field to the detector. The explicit form of $F_\phi$
will be discussed in the next section. 

The optimal strategy to detect a stochastic background requires the 
cross-correlation between the output of (at least) two detectors
\cite{3}, with  uncorrelated noises $n_i(t)\,,i=1,2$. 
Given the two  outputs over a total observation time
$T$,
 \be
s_i(t)=h^i_{\phi}(t)+n_i(t)\,,\quad i=1,2, 
\ee
one constructs a `signal' $S$,  
\be
S=\int_{-T/2}^{T/2}dt\,dt's_1(t)s_2(t')Q(t-t')\,,
\ee
where $Q(t-t')$ is a suitable filter function, usually chosen to
optimise the signal-to-noise ratio:
\be
SNR= \langle S\rangle/ \Delta S,
\ee
where $\Delta S^2= \langle S^2\rangle-\langle S\rangle^2 $ is the
variance of $S$. In our case, we can compute the mean value $\langle
S\rangle$ by using the expansion (\ref{26}) in momentum space, the
statistical independence of the two noises (i.e.  
$\langle n_1(t) n_2(t')\rangle=0$),  and the fact that the noise 
and the strain are uncorrelated (i.e. $\langle n_i(t)
h^i_\phi(t')\rangle=0$). By assuming that the observation $T$ is much
larger than the typical intervals $t-t'$ for which $Q \not=0$, we obtain 
\be
\langle S\rangle=\frac{H^2_0}{5\pi^2}\,T\,\mbox{Re}\,
\left\{\int_0^{\infty}dp\frac{\tilde{Q}(E(p))
\Omega(p)\gamma(p)}{pE^2(p)} \right\}\,,
\label{SNR}
\ee
where 
\be
\tilde{Q}(E(p))=\int_{-\infty}^{\infty} d\tau e^{2\pi\,i(E(p)\tau}Q(\tau)\,,
\ee
and 
\be
\gamma(p)=\frac{15}{4\pi}\,\int d^2\hat{n}
e^{2\pi i p\hat{n}\cdot(\vec{x}_{D1}-\vec{x}_{D2})}
F^{1}_{\phi}(\hat{n})F^{2}_{\phi}(\hat{n}) 
\label{gamma}
\ee
is the so-called overlap reduction function \cite{3}, which 
depends on the relative orientation and location of the two detectors. 
In~(\ref{gamma}) the normalisation constant 
has been chosen so that -- in the massless case -- one
obtains $\gamma(p)=1$ for two  coincident and coaligned
interferometers. 

Switching to the frequency domain ($E=f, dp/df =f/p$),  the mean value
of $S$ can be  written as 
\be
\langle S\rangle=\frac{H^2_0}{5\pi^2}\,T\,\mbox{Re}
\left\{\int_0^{\infty}df\frac{\theta(f-\tilde{m})\,
\tilde{Q}(f)\Omega(\sqrt{f^2-\tilde{m}^2})\gamma(\sqrt{f^2-
\tilde{m}^2})} {(f^2-\tilde{m}^2)f}
\right\}\,,
\label{SNR1}
\ee
where $\tilde{m}=m/2\pi$, and $\theta$ is the Heaviside step function. 
To compute the variance we will assume that for each detector
the noise is much larger in magnitude than the strain induced  by the
scalar wave (i.e. $n_i(t)\gg h^i_{\phi}(t)$). One obtains~\cite{3}
 \be
\Delta S^2\simeq <S^2>\simeq
\frac{T}{2}\,\int_{0}^{\infty}df P_1(|f|)P_2(|f|)|\tilde{Q}(f)|^2\, , 
\ee
where $P_i(|f|)$ is the one-sided noise power spectral density of the $i$-th 
detector, defined by 
\be
\langle n_i(t) n_j(t')\rangle=\frac{\delta_{ij}}{2}\,
\int_{-\infty}^{\infty} df e^{2\pi\,if(t-t')}P_i(|f|)\,.
\ee
Introducing the following positive semi-definite inner product 
in the frequency domain, 
\be
(a,b)\doteq\mbox{Re}\left\{\int^{\infty}_{0}a(f)b(f)P_1(f)P_2(f)
\right\}\, ,
\ee
the signal-to noise ratio can be written as 
\be
(SNR)^2=2\,T\,\left(\frac{H^2_0}{5\pi^2}\right)^2\,
\left[\frac{(\tilde{Q},A)^2}{(\tilde{Q},\tilde{Q})}\right], 
\ee
where 
\be
A=\frac{\theta(f-\tilde{m})\Omega(\sqrt{f^2-\tilde{m}^2})
\gamma(\sqrt{f^2-\tilde{m}^2})}{(f^2-\tilde{m}^2)f\,P_1(|f|)P_2(|f|)}\,.
\ee
The above ratio is maximal if $\tilde{Q}$ and $A$ are
proportional, i.e. $\tilde{Q}=\lambda\,A$. With this optimal choice the 
SNR reads 
\be
SNR=
\left(\frac{H^2_0}{5\pi^2}\right)\,\sqrt{2\,T \,{\cal I}}, 
\ee
where  
\be
{\cal I}=\int_{0}^{\infty}dp
\frac{\Omega^2(p)\,\gamma^2(p)}
{P_1(\sqrt{p^2+\tilde{m}^2})\,P_2(\sqrt{p^2+\tilde{m}^2})
(p^2+\tilde{m}^2)^{3/2}\,p^3}. 
\label{221}
\ee

For $\gamma (p) = (df/dp) \tilde \gamma (f)$ our result
reduces to expression for the SNR already deduced in \cite{13} (modulo a
different normalisation of the overlap function). 
The scalar nature of the background is encoded 
into $\gamma(p)$, and will be discussed in the next Section. 

\section{Antenna patterns and overlap reduction function}
\label{S3}

Given the spectrum, and the noise power spectral densities of 
the two detectors, the computation of the signal-to-noise ratio 
requires now the explicit expression of the overlap
reduction function  $\gamma(p)$ for a pair of gravitational antennas.
As already pointed out in the Introduction, we shall restrict our 
analysis to interferometric detectors. We will also consider 
the interaction of the scalar waves with the differential
mode of the interferometer (see e.g.~\cite{10}), 
which is described  by the following symmetric,  trace-free
tensor\cite{14}:  
\be
D_{ab}=\frac{1}{2}\,(\hat{u}_a\hat{u}_b-\hat{v}_a\hat{v}_b),
~~~~ ~~~a,b= 1,2,3, 
\label{tens_pa}
\ee
where $\hat{u}_a,\hat{v}_a$ are two unit vectors 
pointing in the directions of the arms of the interferometer. We will
therefore neglect the interaction with the common mode, which is
expected  to be much more noisy. 
Note that in this paper we are  not interested
in distinguishing a scalar signal from a tensor one, but only in estimating
the level of the signal eventually induced by a background of scalar waves. 

The computation of $\gamma(p)$ requires knowledge of the
antenna pattern (\ref{27}), which describes the induced strain and
takes into account both the polarisation of the wave and the
geometrical configuration of the detector (parametrised by $D_{ab}$).
As far as the strain induced by a scalar wave is concerned, 
there are two possible contributions~\cite{8}:
one corresponds to the direct interaction of the detector with the scalar
field, while the other is due to the indirect interaction with 
the scalar component of the metric fluctuations induced by the scalar field 
itself. 

Indeed, in a general scalar-tensor theory in which
the matter fields are non-universally and 
non-minimally coupled to the scalar field (for instance, gravi-dilaton
interactions in a superstring theory context \cite{15}), 
the macroscopic bodies are characterised by a composition-dependent
scalar charge (which cannot be eliminated by an appropriate choice of
the frame, like the Jordan frame of conventional Brans-Dicke models),
and their motion in a scalar-tensor background is in general
non-geodesic \cite{8}. Taking into account also the direct coupling of a
test  mass to the gradients of the scalar background field, it follows that
the standard equation of geodesic deviation (which is the main equation 
for deriving the response of a gravitational detector) is 
generalised as follows \cite{8}: 
\be
{D^2 \xi^{\mu}\over D\tau^2} +R_{\alpha\beta\nu}\,^\mu u^\beta u^\nu
\xi^\alpha +q \xi^\alpha \nabla_\alpha \nabla^\mu \phi=0. 
\label{32}
\ee
Here $\xi^\mu$ is the spacelike vector connecting two nearby
(non-geodesic) trajectories, and $q$ is the scalar charge per unit of
gravitational mass, representing the relative strength of scalar interaction 
with respect to ordinary tensor-type interactions \cite{8}. 

The small, non-relativistic oscillations of a test mass, mechanically
equivalent to the detector, are thus described by the equation:
\be
\ddot \xi^a=-\xi^b \left(R_{b00}\,^a +
q \partial_b\partial^a \phi\right). 
\label{33}
\ee
The two contributions to the strain come from the gradients of the
scalar field $\phi$, and from the gradients of the scalar component
$\psi$ of the metric fluctuations (induced by $\phi$), covariantly
represented by the Riemann tensor. The two fields $\phi$ and $\psi$ are
in principle different, but not independent, being related by a set of
coupled differential equations (which are model-dependent). 
The antenna pattern $F_\psi(\hat n)$,
associated to the (indirect) Riemannian part of the strain, has been
computed in \cite{10} for both massless and massive scalar fields. 
We shall see  that, for a traceless detector response tensor, 
the function $F_\psi(\hat n)$ is also proportional to the
antenna pattern $F_\phi(\hat n)$ associated to the direct, non-geodesic
part of the strain. 

Introducing the transverse and longitudinal
projectors with respect to the direction of propagation of the scalar 
wave, 
\be
T_{ab}=(\delta_{ab}-\hat{n}_a\,\hat{n}_b), ~~~~~
L_{ab}=\hat{n}_a\,\hat{n}_b \,,
\ee
the indirect, Riemannian contribution to the scalar pattern function
becomes \cite{10}: 
\be
F_\psi(\hat n)= D^{ab}e_{ab}(\psi)=D^{ab}\left(T_{ab}+
\frac{\tilde{m}^2}{E^2}\,L_{ab}\right).
\label{35}
\ee
Using eq.~(\ref{33}) and the mode expansion for the scalar field, the 
antenna pattern for the direct, non-geodesic coupling is:  
\be
F_\phi(\hat n)= q D^{ab}e_{ab}(\phi)=q D^{ab}\frac{p^2}{E^2}
\,L_{ab}.
\label{36}
\ee
Since $T_{ab}= \delta_{ab}- L_{ab}$, and Tr $D=0$, it follows that 
\be
F_\phi(\hat n)=-qF_\psi(\hat n)=-q \frac{p^2}{E^2}F^0_\psi(\hat n),
\label{37}
\ee
where $F^0_\psi$ is the antenna pattern corresponding to a massless
scalar wave\cite{10}. The overlap reduction
function of two interferometers, directly interacting 
through a charge $q_i$ with a scalar field, can thus be written as   
\be
\gamma(p)=q_1q_2\left(p\over E\right)^4\,\gamma_0(p)\,,
\label{gamma_rel}
\ee
where $\gamma_0(p)$ is the overlap reduction function 
for the geodesic interaction with a massless scalar field~\cite{10}. 
Using Eq.~(\ref{221}) and~(\ref{gamma_rel}), the signal-to-noise ratio is 
finally given by: 
\be
SNR=q_1q_2\,
\left(\frac{H^2_0}{5\pi^2}\right)\, \left[ 2 T 
\int_{0}^{\infty}dp
\frac{p^5\,\Omega^2(p)\,\gamma_0^2(p)}
{P_1(\sqrt{p^2+\tilde{m}^2})\,P_2(\sqrt{p^2+\tilde{m}^2})
(p^2+\tilde{m}^2)^{11/2}}\right]^{1/2}. 
\label{SNR_fin}
\ee

Note that this expression, with $q_i=1$, is also valid to estimate the
signal indirectly induced in the interferometers by the scalar metric 
fluctuations through their
usual coupling to the Riemann tensor, provided $\Omega(p)$  refers to
the associated  
spectrum of scalar metric fluctuations. In the following
we shall therefore use eq. (\ref{SNR_fin}) setting $q_1=q_2=1$ if the
dominant signal comes indirectly 
from the stochastic background of scalar metric 
fluctuations $\Omega_\psi(p)$ induced by the scalar field,  
and setting instead $q_i<1$ (according to the experimental constraints) 
if the dominant signal is directly due to the  stochastic background 
$\Omega_\phi(p)$ of massive scalar waves. 

\section{Examples}
\label{S4}

We will now apply the results of the previous sections to estimate the
signal induced by a stochastic background due to a massive scalar field 
on the two LIGO interferometers; 
we will consider the two detectors operating   
in the initial (I), enhanced (II) and advanced (III) 
configurations \cite{16}. In particular, for the noise spectral density we 
will use the following analytical fits:
\begin{itemize}
\item LIGO I ~\cite{Sathya}:
\begin{eqnarray}
&&P(f)=\frac{3}{2}\,P_0
\left[\left(\frac{f}{f_0}\right)^{-4}+2+
2\left(\frac{f}{f_0}\right)^2\right], \nonumber \\
&& P_0=10^{-46}\mbox{Hz}^{-1},
\quad\,f_{s}=40\mbox{Hz},
\quad\,f_{0}=200\mbox{Hz}.
\label{nligo1}
\end{eqnarray}
\item LIGO II ~\cite{BenSathya}: 
\begin{eqnarray}
&&P(f)=\frac{P_0}{11}\,
\left[\left(\frac{f}{f_0}\right)^{-9/2}
+\frac{9}{2}\left(1+\left(\frac{f}{f_0}\right)^2\right)\right], 
\nonumber \\
&& P_0=7.9\times10^{-48}\mbox{Hz}^{-1},
\quad\,f_{s}=25\mbox{Hz},
\quad\,f_{0}=110\mbox{Hz}.
\label{nligo2}
\end{eqnarray}
\item LIGO III ~\cite{BenSathya}:
\begin{eqnarray}
&&P(f)=\frac{P_0}{5}\,
\left[\left(\frac{f}{f_0}\right)^{-4}+2+
2\left(\frac{f}{f_0}\right)^2\right], \nonumber \\
\nonumber \\
&& P_0=2.3\times10^{-48}\mbox{Hz}^{-1}, 
\quad\,f_{s}=12\mbox{Hz}, 
\quad\,f_{0}=75\mbox{Hz}. 
\label{nligo3}
\end{eqnarray}
\end{itemize}
In the above equations, $f_{s}$ is a seismic cut-off below which 
the noise spectral density is treated as infinity. 
We shall  also assume that the non-relativistic modes in the 
spectrum are the dominant ones, and their energy density almost 
saturates the critical energy density, namely
\be
\int_0^m d\ln p~\Omega^{\rm non-rel}(p)\simeq 1. 
\label{41}
\ee
We shall analyse, in particular, two examples of spectra.

\subsection{Minimal dilaton background}

The first example is a stochastic background of massive relativistic
dilatons, produced according to some models of  early cosmological
evolution based upon string theory \cite{7}. 
To illustrate the difficulty in detecting such a
background, we will consider here the dilaton spectrum obtained in the
context of a ``minimal" pre-big bang scenario and we shall assume that the
dilaton mass is enough small so that the produced dilatons have not yet
decayed into radiation with a rate $\Gamma \sim m^3/M_P^2$. This
implies $\Gamma \laq H_0$, i.e $ m \laq 100$ MeV. 

Even if the mass is negligible at the beginning of the radiation era, the
proper momentum $p=k/a(t)$ is red-shifted with respect to the rest
mass because of the cosmological expansion, and all the modes in the 
dilaton spectrum tend to become non-relativistic. As a consequence, the
present dilaton spectrum has in general at least three branches,
corresponding to: {\em i)} relativistic modes, with $p>m$;  {\em ii)}
non-relativistic modes with $p_m<p<m$,  which became non-relativistic
inside the horizon;  {\em iii)} non-relativistic modes with $p<p_m$, 
which became non-relativistic outside the horizon. The separation
scale $p_m$ corresponds to a mode that became non-relativistic,
$p\simeq m$, just at the time  of horizon crossing, $p\simeq H$, and
is given by \cite{17} $p_m =p_1(m/H_1)^{1/2}$, where $H_1$ is the
Hubble scale at the transition between the inflationary phase and the 
radiation phase,  and $p_1 \simeq
H_1a_1/a \simeq (H_1/M_P)^{1/2} 10^{11}$ Hz is the high frequency
cut-off of the spectrum. 

The complete dilaton spectrum, at the present
time $t_0$, can thus be parametrised as follows \cite{18}:
\be
\Omega(p)\simeq\left\{
\begin{array}{rcl}
 &&   \left(H_1\over M_p\right)^2 
\Omega_\gamma (t) \left(p\over p_1\right)^\delta, 
~~~~~~~~~~~~~~~~~m< p <p_1, 
\\[4pt]
&&   \left(H_1\over M_p\right)^2 
\left(m^2\over H_1H_{\rm eq}\right)^{1/2} \left(p\over
p_1\right)^{\delta-1},  ~~~~~~ p_m< p <m, 
\\[4pt]
&&   \left(H_1\over M_p\right)^2 
\left(m\over H_{\rm eq}\right)^{1/2} \left(p\over
p_1\right)^{\delta},  ~~~~~~~~~~~~ p< p_m, 
\end{array}
\right . \label{42}
\ee
where $\Omega_{\gamma}\sim 10^{-4}$ is the present photon energy
density (in critical units),   
$H_{\rm eq}\sim 10^{-55}M_{P}$ is the Hubble scale at the epoch of
matter-radiation equilibrium, and $\delta\ge 0$ is a slope parameter,
depending on the kinematics of the pre-big-bang phase \cite{7}
($\delta =3$ for the low energy dilaton-driven phase, while $\delta <3$
for modes crossing the horizon during the high curvature phase). 

Let us recall now that if the dilaton is strongly coupled to macroscopic
matter with a composition-dependent charge $q \gaq 1$, as suggested
by loop corrections to the string effective action \cite{19}, then the
mass has to be large enough, $m \gaq 10^{-4}$ eV $\sim 10^{11}$ Hz, to
be compatible with known phenomenological bounds \cite{20}. In that
case,  the mass is too far from the sensitivity range of a typical
interferometric antenna, and no signal is detectable, since the spectral
noises $P_i$ are dominated by the mass, and tend to infinity.  

To discuss a possible detection, let us thus 
consider the alternative possibility of (almost) universal string loop
corrections discussed in \cite{11}, which is compatible with very light
(but weakly coupled) dilatons, 
and let us assume the most favourable
situation in which: {\em i)} the  frequency corresponding to the 
dilaton mass lies within the
sensitivity band of the interferometers; {\em ii)} $\delta <1$, and  its
value  is enough small so that the spectrum 
is dominated by the non-relativistic branch;
{\em iii)} the spectrum saturates the critical density bound. In such a
case,  using the explicit form (\ref{42}), it is easy to show that the main
contribution to the integral (\ref{41}) comes from the low frequency
band $p<p_m$, 
and that the mass and the spectral index are related through the 
following  equation: 
\be 
(H_1/M_P)^{2}\,(m/H_{\rm eq})^{1/2}\,(m/H_{1})^{\delta/2}=\delta. 
\label{43}
\ee

With such a small mass, however, the dilatonic interaction is long-range 
and  the direct coupling to matter is strongly suppressed ($q \ll 1$
according to \cite{20}): the SNR turns out to be negligible, as shown by an
explicit computation. Let's  therefore make the further assumption
that the dilaton background induces scalar metric fluctuations with a
spectrum which has the same amplitude and the same frequency
distribution as the dilaton spectrum (\ref{42}) (this is indeed what
happens for relativistic dilatons, at the tree-level, thanks
to the gravi-dilaton mixing of the string effective action \cite{7}). Such
a background of scalar metric fluctuations would be gravitational
coupled to the detectors with $q=1$. 

We have computed the signal-to-noise ratio
(\ref{SNR_fin}) obtained by cross-correlating 
the two LIGO detectors with $q_1=q_2=1$, using the 
background~(\ref{42}) with $H_1$ fixed around
the string scale $M_S\simeq H_1 =10^{-1}M_P$, and
the slope $\delta$ determined by the condition (\ref{43}). 
We have assumed $T=10^7$sec, and the overlap function $\gamma_0(p)$ 
has been computed using the coordinates of the two LIGO sites 
(see e.g. \cite{Allen_coo}). 

The results of this computation are summarised in Fig. 1, where 
the SNR is plotted as a function of the dilaton mass. The scalar signal 
would be detectable at a good
confidence level by LIGO III, for a dilaton mass in the range $ 20 -
300$ Hz.  Unfortunately, however, such values of $m$ compatible with
detection would imply for the metric fluctuation spectrum a maximum 
intensity $\Omega \sim 1$ around a peak scale $p_m \sim 10^{-10} -
10^{-9}$ Hz.  It follows from eq. (\ref{42}) that $\Omega \gaq
10^{-2}$ at $10^{-8}$ Hz, in sharp contradiction with the bound on metric 
fluctuations obtained from pulsar-timing data \cite{21} (which requires 
$\Omega \laq 10^{-8}$ at $10^{-8}$ Hz). Imposing the pulsar-timing 
constraint  on the induced spectrum of scalar metric 
fluctuations it turns out that the signal is not detectable. 

\begin{figure}[t]
\begin{center}
{\epsfig{file=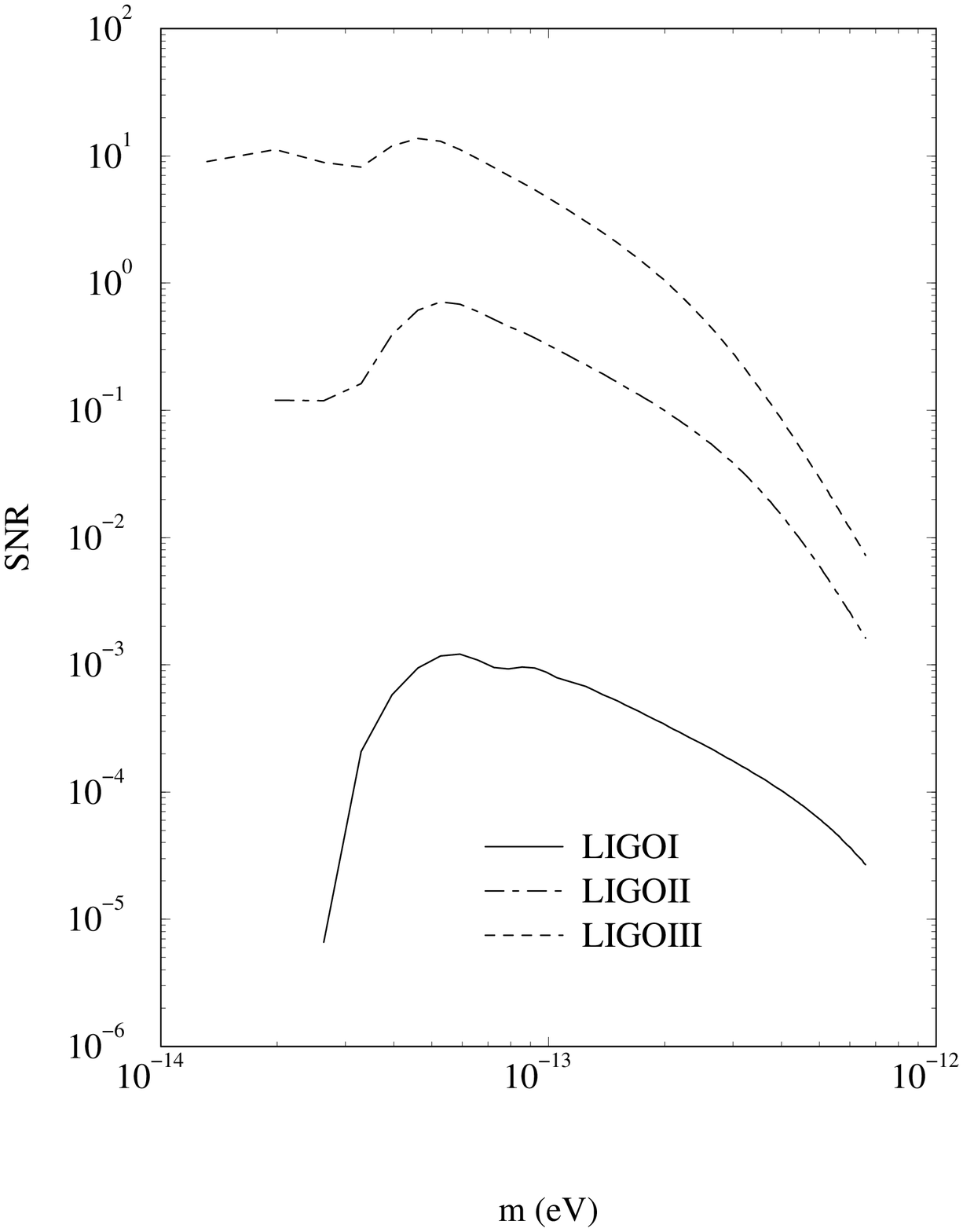,
angle=0,width=5.4in,bbllx=25pt,bblly=50pt,bburx=590pt,bbury=740pt
}}
\end{center}
\label{f_uno}
\vskip 5mm
\caption{\sl The signal-to-noise ratio (\ref{SNR_fin}) for the
cross-correlation of the two LIGO interferometers, coupled with
maximal strength ($q_1=q_2=1$) to the scalar spectrum
(\ref{42}), which is normalised so as to saturate the critical density at
$p=p_m$. The integration time is $T=10^7$sec. The signal is possibly 
non-negligible only if the mass is in the sensitivity band of the
gravitational detectors.  It must be noted, however, that the
extrapolation of the spectrum to much lower scales is not compatible
with the bounds obtained from pulsar timing data.}
\end{figure}

\subsection{Non-relativistic scalar dark matter}

The difficulties in detecting the minimal dilaton background discussed in 
the previous section are mainly due to the fact
that the spectrum (\ref{42}) is characterised by a peak located at a
scale $p=p_m$, much lower than the mass scale $p=m$. However, if the peak  
of the  spectrum is not too far from the mass scale, 
the signal could be detected
even through the direct interaction of the scalar background with the
interferometers, with an effective coupling compatible with the existing 
phenomenological bounds. 

To illustrate this possibility we shall therefore consider a stochastic
background of  scalar massive particles whose energy density represents
a significant fraction of the critical energy density.  We will 
assume that the spectrum is dominated by non-relativistic modes,  with
a peak  for $p\sim m$. For the sake of simplicity we use the following 
single-power law for the spectrum: 
\be
\Omega(p)=\delta ~\Omega_s\,\left(\frac{p}{m}\right)^{\delta}\,,\quad
p\leq m\,,\,\,~~~\delta\geq 0 ,
\label{44}
\ee
where the parameter $\Omega_s$ is the total energy density  (in units of
the critical energy density) of the spectrum, integrated over all modes.
Since we are interested in  a mass for the scalar field corresponding to a
frequency in the range  $10 - 10^3$ Hz, we shall
assume that the scalar charge $q$ is consequently suppressed: taking 
into account all existing phenomenological bounds \cite{20}, it turns out 
that in this frequency range the allowed coupling can be parametrised
as  \be
\log q^2 \laq -7 +\log (m/10~ {\rm Hz}), ~~~~
10 ~{\rm Hz} \laq m \laq 1~  {\rm kHz} 
\label{45}
\ee
for universal scalar interactions, and 
\be
\log q^2 \laq -8 +\log (m/10~ {\rm Hz}), ~~~~
10 ~{\rm Hz} \laq m \laq 1~  {\rm kHz} 
\label{46}
\ee
for composition-dependent interactions. 

We have computed the signal-to-noise ratio,  obtained by
cross-correlating the  two LIGO detectors, for the spectrum~(\ref{44}) 
assuming  a nearly critical energy density ($\Omega_s=1$) and the 
maximal value of $q$ allowed by the bound (\ref{45}). Fig.~2 shows the
contour plots  of the SNR for $T=10^{7}$sec, as a function of
$\log_{10}(m)$ and of the  spectral index $\delta$,  
for the initial (I), enhanced (II) and advanced (III) LIGO configurations. 
We have considered the case $\delta \leq 3$, typical of  string cosmology
models \cite{7}. Already with the enhanced LIGO configuration, the signal
produced by  this class of backgrounds may have a significant 
signal-to-noise ratio in the relevant mass range. 
Similar results can be obtained with a composition dependent  scalar
charge,  provided the values of SNR are rescaled by a factor $10^{-1}$,
according to eq. (\ref{46}). 

Fig~3 shows the values of $q^2 \Omega_s$
corresponding to SNR=1 for the enhanced and advanced LIGO
configurations, as a function of the scalar mass, using the spectrum
(\ref{44}) with $\delta=1$ (which is a typical value for 
the slope of the tensor perturbations spectrum 
predicted in quintessential inflation models~\cite{27}). 
The interesting result is that, in the absence of any signal,  an upper limit
on $q^2 \Omega_s$  would imply an indirect bound on the scalar
charge stronger than the direct bounds existing at present 
\cite{20}, provided the expected energy density of the scalar background
corresponds to a non-negligible fraction of critical energy density. 

\begin{figure}[t]
\begin{center}
\includegraphics[width=8cm,height=8cm]{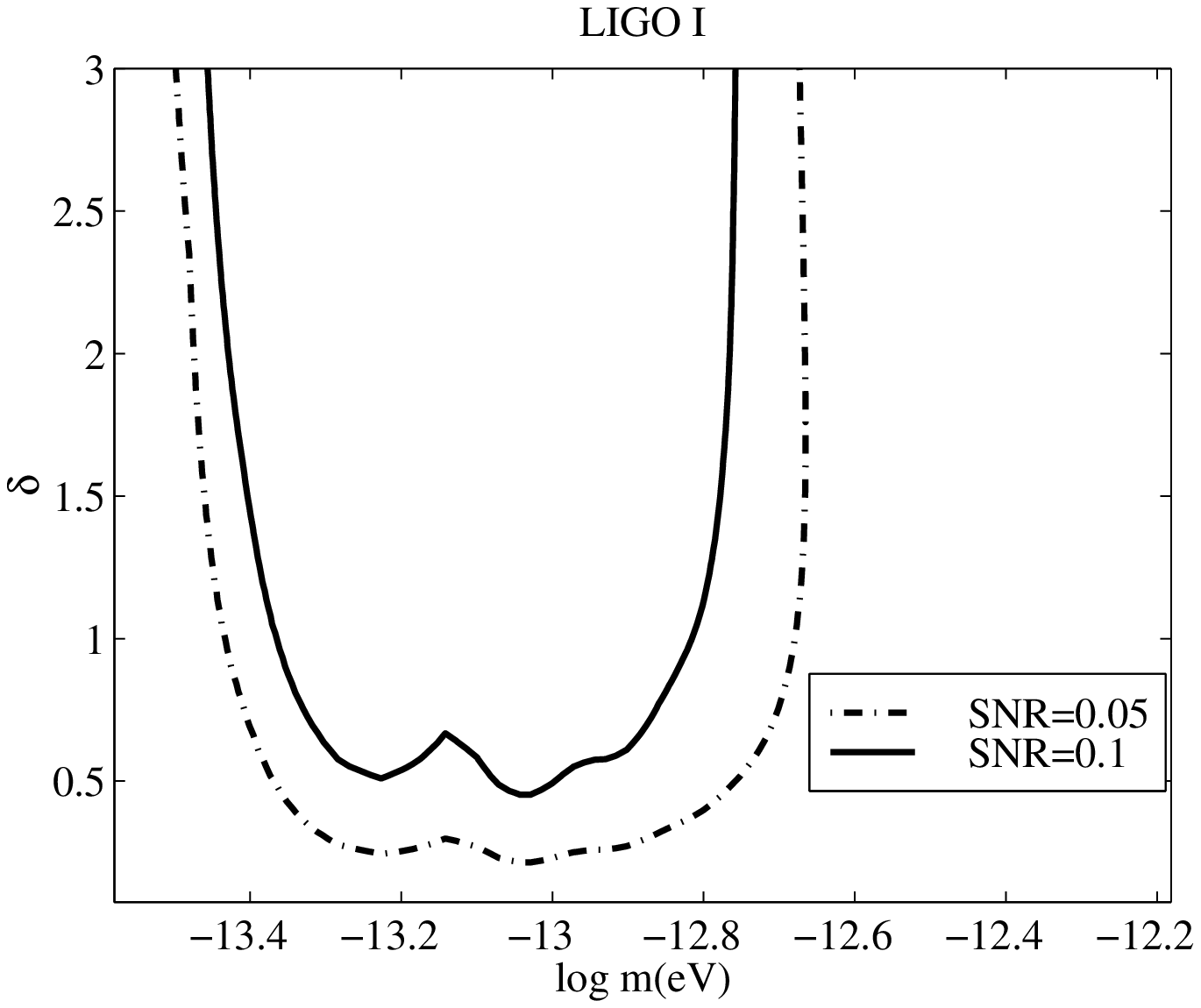}
\includegraphics[width=8cm,height=8cm]{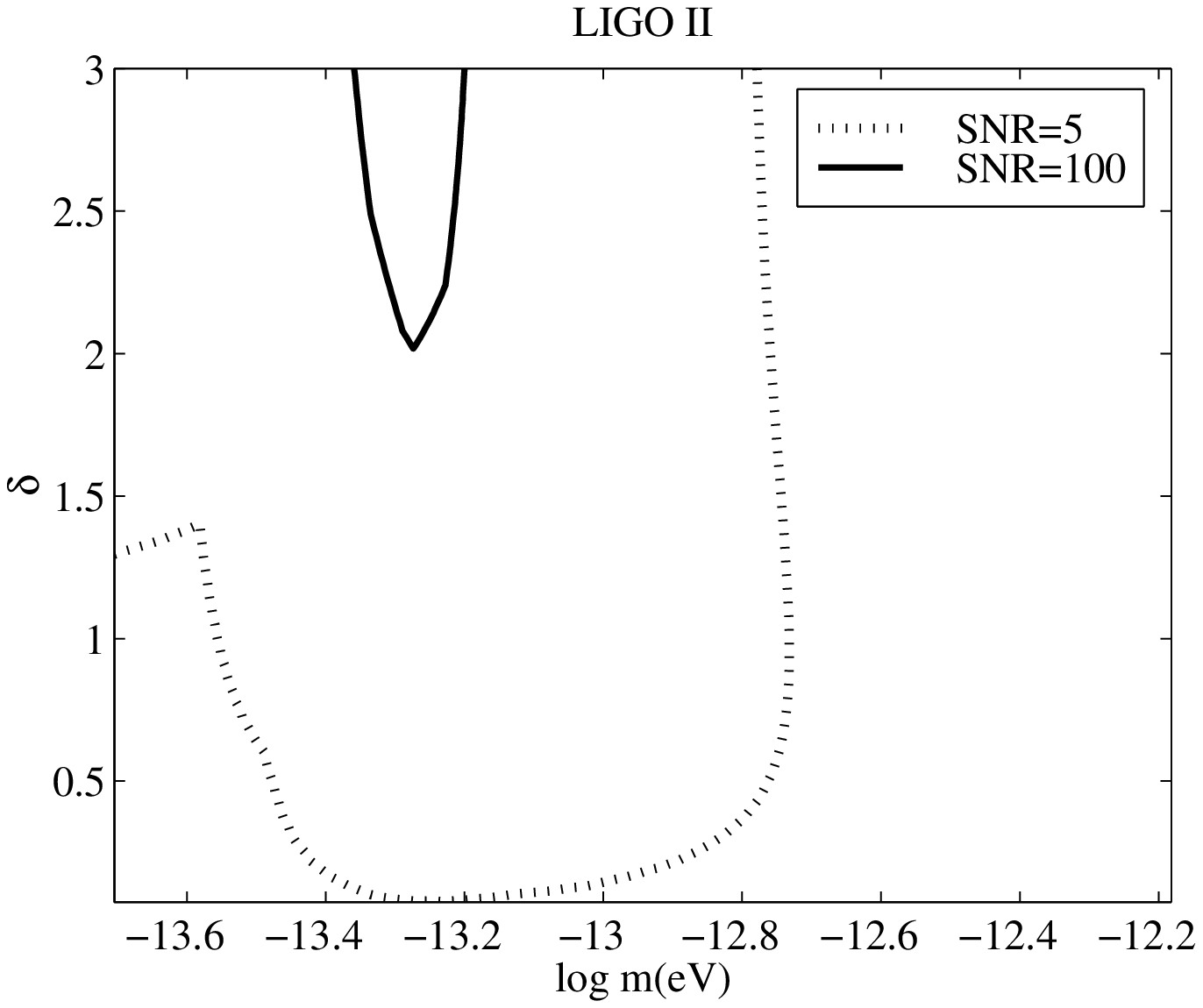}
\includegraphics[width=8cm,height=8cm]{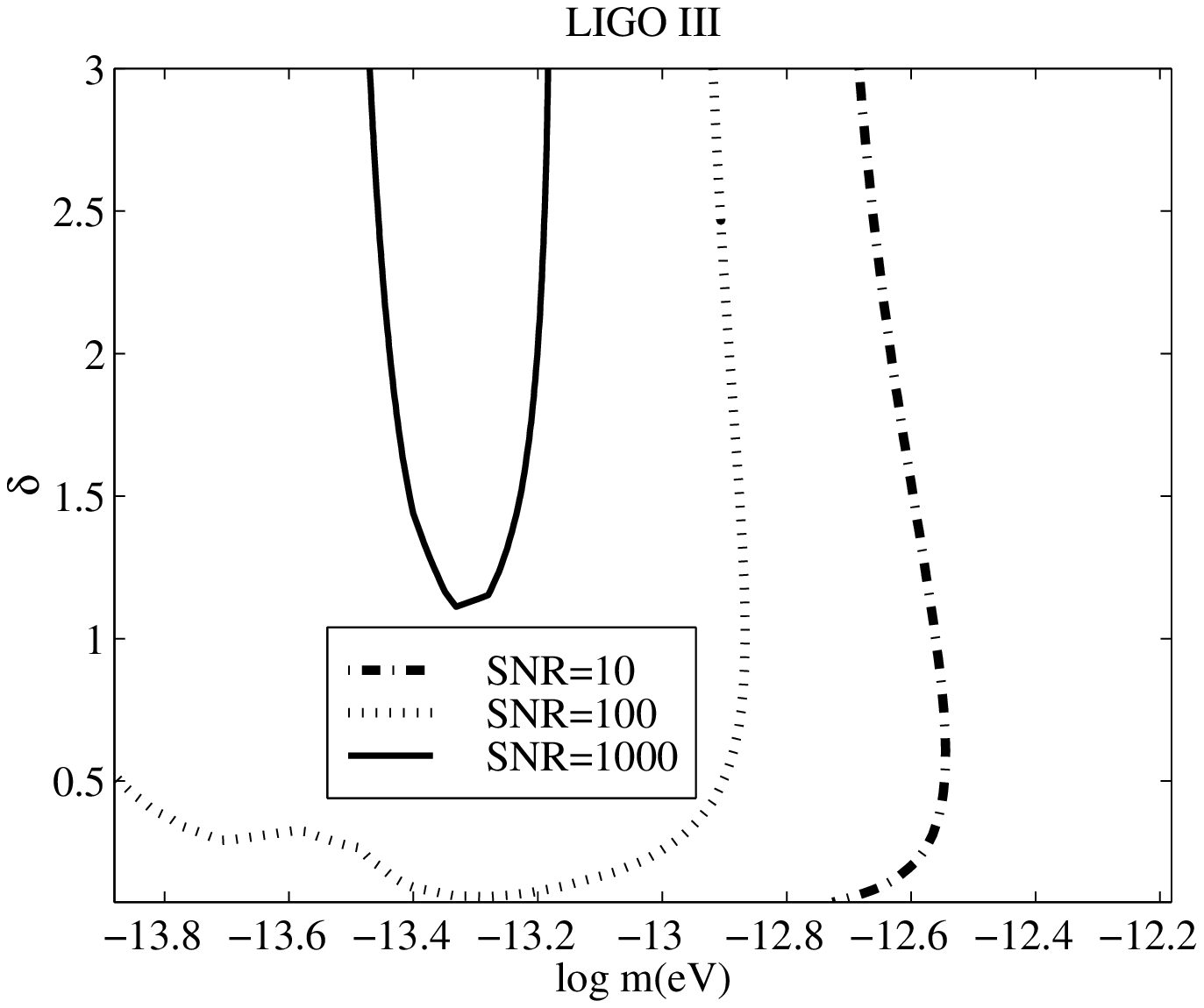}
\end{center}
\label{f_due}
\caption{\sl Contour-plots in the plane $\{m,\delta \}$ for the SNR of
eq. (\ref{SNR_fin}), computed for the scalar spectrum (\ref{44}), and
relative to the cross-correlation of a pair of 
LIGO interferometers in the initial (I), enhanced (II) and advanced (III) 
configurations. We have used $T=10^{7}$sec, $\Omega_s=1$, 
and the maximal allowed value of $q$ according to eq. (\ref{45}). 
The region inside the contour corresponds to a value of SNR bigger 
than the value on the border.}
\end{figure}

\begin{figure}[t]
\begin{center}
{\epsfig{file=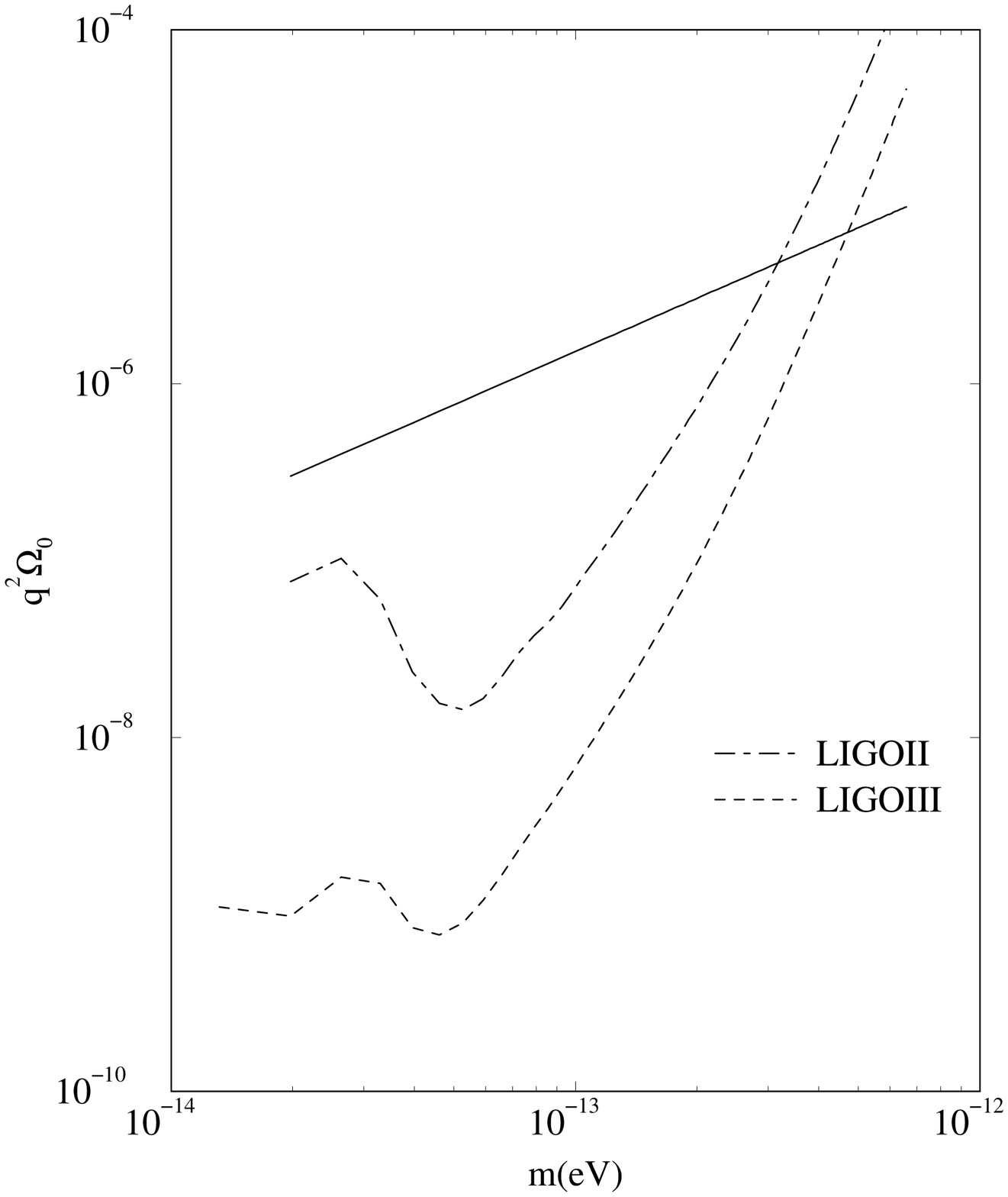,
angle=0,width=5.4in,bbllx=25pt,bblly=50pt,bburx=590pt,bbury=740pt
}}
\end{center}
\label{f_tre}
\caption{\sl Allowed values of $q^2 \Omega_s$, as a function of the
mass of the scalar  field, and corresponding to  SNR = 1,  computed for
the scalar spectrum (\ref{44}) and for the two LIGO
detectors in the enhanced and advanced  configurations. For
comparison, we have also plotted the maximal value of $q^2$  according
to eq.~(\ref{45}) (solid line). We have used a total observation time  
$T=10^{7}$sec, and a spectral index  $\delta=1$.} 
\end{figure}

\section{Conclusions}
\label{S5}

In this paper we have discussed the possible detection of a relic
background of massive, non-relativistic scalar particles, through the
cross-correlation of two interferometric gravitational antennas. 

Using previous results \cite{8,10,13}, we have computed the
signal-to-noise ratio for a stochastic background of massive scalar
waves, taking into account both the scalar and gravitational charge of 
the detector, i.e. including in the differential mode of the 
interferometer both the direct coupling to the scalar field spectrum and 
the coupling to the induced  spectrum of scalar metric fluctuations. 

We have computed the signal-to-noise ratio obtained by cross-correlating the 
two LIGO interferometers in the initial, enhanced and advanced configurations,
for two particular examples: relic dilatons produced in the context of
string cosmology models, and a generic scalar component of dark
matter. 

If the total energy density of the scalar background is
dominated by non-relativistic modes and amounts to a significant
fraction of the critical energy density, we have found that the induced
signal is certainly non-negligible for the LIGO enhanced and advanced
configurations, provided both the mass and the peak of the spectrum
lie within the sensitivity band of the detectors,  i.e. from $10$ Hz to
$1$ k Hz. These results hold even if the direct coupling to the detector is
strongly suppressed, according the existing bounds on long range
scalar interactions. 

It is worth pointing out that the differential mode of an
interferometer is particularly unfavoured for the detection of 
non-relativistic scalar waves: since the detector response tensor 
is traceless, the corresponding pattern function, for non-relativistic
modes, is suppressed by a factor $(p/E)^2 \simeq (p/m)^2 \ll 1$. This
is the reason why, to obtain a detectable signal, we have restricted our
discussion to spectra peaked around $p=m$. However, for different
detectors (for instance, resonant spheres) the detector response tensor 
with respect a scalar wave is not traceless, and the suppression factor 
is absent. It would be interesting to investigate 
the detectability of a more general class of stochastic backgrounds 
of massive scalar particles (i.e. where the peak and the mass are not 
in the same frequency range) through resonant spherical detectors. 

The main conclusion of our work, confirming the results of previous
studies,  is that  a stochastic background of light but non-relativistic
scalar particles,  which provides a significant fraction of critical energy
density, could induce a non-negligible signal in future enhanced
interferometric detectors  (in competition with the signal of a possible
stochastic background of  gravitational waves), provided the frequency
corresponding to the mass of  the scalar field lies within the detector
sensitivity band.  We believe that this possibility should be taken into
account  in the data analysis development of second-generation
interferometric  detectors.

\acknowledgments
It is a pleasure to thank Bruce Allen, Bruce Bassett, Danilo Babusci, Francesco
Fucito, Massimo Giovannini, Michele Maggiore and  Gabriele Veneziano
for helpful comments and/or discussions. We also thank
B.~Sathyaprakash for having  provided an analytical fit of the power 
 spectral density  for the noise of the initial LIGO configuration.

\end{document}